\documentstyle[twoside,fleqn,espcrc2,psfig]{article}

\newcommand{\la}{\langle}
\newcommand{\ra}{\rangle}

\title{
       Nucleon decay matrix elements with the Wilson quark action:\\ an update
       \thanks{presented by N.~Tsutsui}
      }

\author{
        JLQCD Collaboration:
        S.~Aoki
          \address{Institute of Physics, University of Tsukuba,
          Tsukuba, Ibaraki 305-8571, Japan},
        M.~Fukugita
          \address{Institute for Cosmic Ray Research,
          University of Tokyo, Tanashi, Tokyo 188-8502, Japan},
        S.~Hashimoto
          \address{High Energy Accelerator Research
          Organization (KEK), Tsukuba, Ibaraki 305-0801, Japan},
        K-I.~Ishikawa$^{\rm c}$,
        N.~Ishizuka$^{\rm a,}$
          \address{Center for Computational Physics,
          University of Tsukuba, Tsukuba, Ibaraki 305-8577, Japan},\\
        Y.~Iwasaki$^{\rm a,d}$,
        K.~Kanaya$^{\rm a,d}$,
        T.~Kaneda$^{\rm a}$,
        S.~Kaya$^{\rm c}$,
        Y.~Kuramashi$^{\rm c}$,
        M.~Okawa$^{\rm c}$,
        T.~Onogi
          \address{Department of Physics, Hiroshima
          University, Higashi-Hiroshima, Hiroshima 739-8526, Japan},\\
        S.~Tominaga$^{\rm c}$,
        N.~Tsutsui$^{\rm e}$,
        A.~Ukawa$^{\rm a,d}$,
        N.~Yamada$^{\rm e}$,
        T.~Yoshi\'e$^{\rm a,d}$
       }

\begin{document}
%%%%%%%%%%%%%%%%%%%%%%%%%%%%%%%%%%%%%%%%%%%%%%%%%%%%%%%%%%%%%%%%%%%%%%
\begin{abstract}
We present preliminary results of a new lattice computation of hadronic
matrix elements of baryon number violating operators which appear in the
low-energy effective Lagrangian of (SUSY-)Grand Unified Theories.  The
contribution of irrelevant form factor which has caused an underestimate
of the matrix elements in previous studies is subtracted in this
calculation.  Our results are 2$\sim$4 times larger than the most
conservative values often employed in phenomenological analyses of
nucleon decay with specific GUT models.
\end{abstract}

\maketitle
%%%%%%%%%%%%%%%%%%%%%%%%%%%%%%%%%%%%%%%%%%%%%%%%%%%%%%%%%%%%%%%%%%%%%%
\section{Introduction}
Nucleon decay is an important consequence of Grand Unified Theories
(GUTs), which can give strong constraints on such theories.
Unfortunately, theoretical estimations of the lifetime and branching
ratios suffer from large uncertainties.  One of the sources of the
uncertainties is the value of hadronic matrix elements from 
the nucleon to pseudo scalar meson through baryon number violating 
operators which appear in the low-energy
effective Lagrangian of GUTs.

Conventionally these matrix elements are estimated by rewriting them 
in terms of the three-quark annihilation amplitude $\la 0 | O | p \ra$ using the
soft-pion theorem.  There have been several lattice calculations for
$\la 0 | O | p \ra$ \cite{Hara86,Bowler88,Gavela89}.  They have
not yielded a definitive result, however.  Moreover, the validity of the 
soft-pion theorem for the nucleon decay process has not been clarified.

A pioneering lattice work to calculate $\la \pi^0 | O | p \ra$ directly
was made in \cite{Gavela89}.  We also made an effort to advance it
in several fronts \cite{JLQCD98}. After this study, we have found that
previous lattice calculations for the matrix elements including ours 
had a subtle point which caused an underestimate of the matrix elements.
The problem is that the previous results contain the contribution of an 
irrelevant form factor,  which vanishes after using the equation of
motion of the out-going lepton and setting its mass to zero.  This
contribution can not be subtracted using our previous data, essentially
because we did not measure all necessary components of the three-point
function.

In this article we report results of a new simulation in which we measure 
all the necessary components
of the three-point correlation function to disentangle the relevant form
factor from the irrelevant one.

As in our previous calculation, we evaluate matrix elements of
all dimension six baryon number violating operators classified according
to $SU(3) \times SU(2) \times U(1)$ \cite{Weinberg79,Wilczek79}, so as
to cover various GUT models and decay processes.  Also the calculations 
are made at physical quark mass and physical momentum in order to
take account of the $N \to K$ process which is expected to be the
dominant mode for SUSY-GUTs.

%%%%%%%%%%%%%%%%%%%%%%%%%%%%%%%%%%%%%%%%%%%%%%%%%%%%%%%%%%%%%%%%%%%%%%
\section{Calculational method}
We wish to calculate the matrix element of a class of three quark 
operators between the nucleon N and a pseudoscalar meson PS. 
Using Lorentz and parity invariance, one finds
that the matrix elements are described by two form factors,
\begin{eqnarray}
&& \la PS | (q_1 C P_{R/L} q_2) P_{R/L} q_3 | N \ra \nonumber\\
&=& A(q^2) P_{R/L} N + B(q^2) P_{R/L} \not{\!q} N,
\end{eqnarray}
where $q_i$ represents a quark field, $q$ is a four momentum of the
out-going lepton and $N$ and $P_{R/L}$ stands for the nucleon spinor
and chiral projection operator, respectively.

Multiplying by the lepton spinor and using the equation of motion, one can
see that the second term is proportional to lepton mass.  We
call $A$ and $B$ as relevant and irrelevant form factor,  respectively,
as the latter does not contribute when the lepton mass is neglected.

To extract the form factors, we form a ratio,
\begin{equation}
\frac{\la J_{PS}(t_{y}) O_{\gamma}(t_{x}) \bar{J}_{N,\gamma^{\prime}}(0) \ra}
{ \la J_{PS}(t_{y}) J_{PS}^{\dagger}(t_{x}) \ra
  \la J_{N}(t_{x}) \bar{J}_{N}(0) \ra }
\sqrt{Z_{PS}} \sqrt{Z_N},
\end{equation}
where $\gamma$ and $\gamma^{\prime}$ are spinor indices, and
$Z_{PS}$ ($Z_N$) is the residue of the two-point function
of PS meson (nucleon).  This ratio has the following asymptotic form,
\begin{equation}
\left(
\begin{array}{cc}
   A + q^0 B                  & 0 \\
   B \vec{q}\cdot\vec{\sigma} & 0 
\end{array}
\right),
\end{equation}
in a $2 \times 2$ block notation.

It is important that the upper component is a linear combination of the two
form factors and the lower component is proportional to the irrelevant
term.  In order to disentangle the relevant form factor from the
irrelevant one, both components are needed, while, in the previous
studies, only the upper component was measured.  It should be noted that
we can not follow this procedure for the case $\vec{q}$=0.

To compare our results with the predictions of the tree-level chiral
Lagrangian \cite{Claudson82,Chadha83}, we also calculate the three-quark
annihilation amplitude,
\begin{eqnarray}
\la 0 | (u C d_R) u_L | p \ra &=& \alpha N_L, \\
\la 0 | (u C d_L) u_L | p \ra &=& \beta N_L,
\end{eqnarray}
which are obtained from the two-point function.

%%%%%%%%%%%%%%%%%%%%%%%%%%%%%%%%%%%%%%%%%%%%%%%%%%%%%%%%%%%%%%%%%%%%%%
\section{Numerical simulation}
Our calculation is carried out in quenched QCD at $\beta$=6.0 with the
Wilson quark action on a $28^2 \times 48 \times 80$ lattice.  We analyze
100 configurations at the hopping parameter $K$=0.15620, 0.15568,
0.15516, 0.15464.  The lattice scale fixed by $m_\rho$=770 MeV in the
chiral limit ($K_c$=0.15714(1)) is $a^{-1}$=2.30(4) GeV, and the point
for the strange quark estimated from $m_K/m_\rho$=0.644 is given by
$K_s$=0.15488(7).

We fix the nucleon source at $t=0$, PS meson sink at $t=29$ and move the
operator between them.  Matrix elements are evaluated at four spatial
momenta $\vec{q}a$=(0,0,0), ($\pi$/14,0,0), (0,$\pi$/14,0),
(0,0,$\pi$/24), injected in the PS meson sink.  As mentioned in the
previous section, it is not possible to disentangle the relevant form
factor from the irrelevant one when $\vec{q}$=0. Therefore, we use two
data with two finite momenta to interpolate to the physical momentum.

The form factor obtained by fitting the ratio of three-point function
and two-point functions depends on the quark mass and the momentum
squared.  We distinguish $u$-$d$ and $s$ quark masses; the
former ($m_{ud}$) is taken to the chiral limit, and the latter ($m_s$) 
interpolated to the
physical $s$ quark mass.  We also interpolate the momentum square to the
physical lepton mass.  To do this, we fit the data with the following
form,
\begin{equation}
c_1 + c_2 \cdot q^2 + c_3 \cdot (q^2)^2 + c_4 \cdot m_{ud} + c_5 \cdot m_s.
\end{equation}

Matrix elements are renormalized, with mixing included, by
tadpole-improved one-loop renormalization factors \cite{Richards87} to
the $\overline{{\rm MS}}$ scheme calculated at the scale $\mu=1/a$.  The
quoted errors are only statistical, estimated by the single
elimination jackknife procedure.

%%%%%%%%%%%%%%%%%%%%%%%%%%%%%%%%%%%%%%%%%%%%%%%%%%%%%%%%%%%%%%%%%%%%%%

\begin{figure}[hbt]
\psfig{file=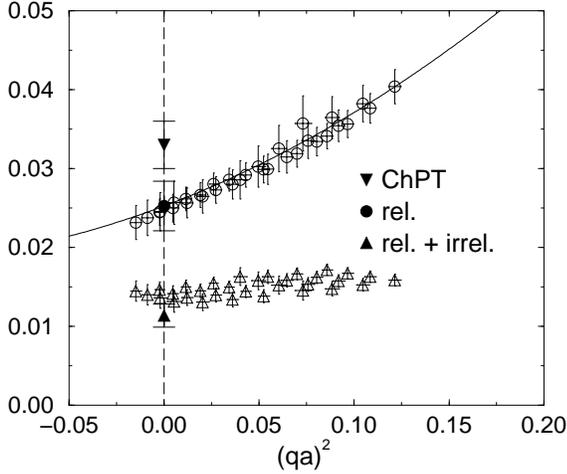,width=7.5cm}
\vspace{-10mm}
%\vspace{9pt}
%\framebox[55mm]{\rule[-21mm]{0mm}{43mm}}
\caption{$q^2$ dependence of the relevant form factor $|A_\pi(q^2)|$
 (circles) and a linear combination of the relevant and irrelevant form
 factors $|A_\pi(q^2)+q^0B_\pi(q^2)|$ (up-triangles), obtained from the
 matrix element $\la \pi^0 | (u_R C d_R)u_L | p \ra$.  The prediction
 from chiral Lagrangian is also shown (down-triangle).}
\label{fig:fit}
\vspace{-8mm}
\end{figure}

\section{Results}
We present a typical $q^2$ dependence of the relevant form factor in
Fig.~\ref{fig:fit}.  We also plot a linear combination of the relevant and
irrelevant form factors $A+q^0B$ and the prediction from tree-level
chiral Lagrangian using $|\alpha|$=0.015(1) GeV$^3$ obtained in our
simulation (we also obtain $|\beta|$=0.014(1) GeV$^3$).  We see
that the value of the relevant form factor is close to that from the chiral
Lagrangian.  As expected from analytical considerations with the chiral
Lagrangian, there is a cancellation between the relevant and irrelevant form
factors.  We consider that this led to the inconsistency between the
direct computation through three-point functions and the indirect
computation using the tree-level chiral Lagrangian reported in previous
lattice studies \cite{Gavela89,JLQCD98}.

\begin{figure}[bt]
\psfig{file=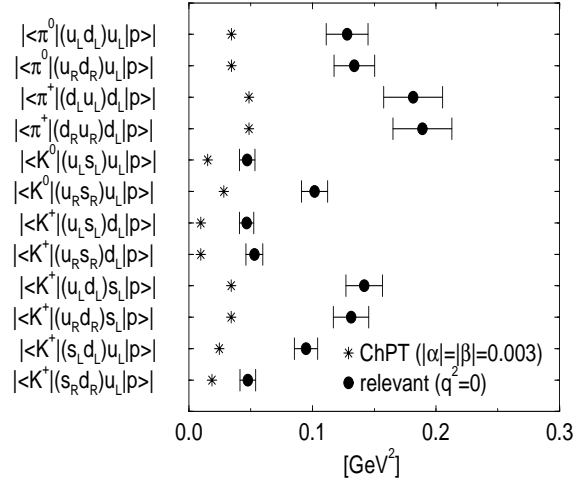,width=7.5cm,height=6.5cm}
\vspace{-10mm}
%\vspace{9pt}
%\framebox[55mm]{\rule[-21mm]{0mm}{43mm}}
\caption{Relevant form factors at physical momentum $q^2=0$ from present
calculation (filled circles) compared with the most conservative
predictions of tree-level chiral Lagrangian (asterisks).}
\label{fig:result}
\vspace{-4mm}
\end{figure}

In Fig.~\ref{fig:result}, we plot our results for the relevant form factors
at $q^2$=0.  To compare our results with the most conservative
estimation of the matrix elements, often employed in phenomenological
analyses of nucleon decay with specific GUT models, we also plot the
prediction from tree-level chiral Lagrangian with a choice of the
parameters $|\alpha|$=$|\beta|$=0.003 GeV$^3$.  
Our results are about 2$\sim$4 times larger than the most conservative
values, implying a stronger constraint on the parameter space of GUT
models.

\vspace*{9pt}
This work is supported by the Supercomputer Project No.45 (FY1999)
of High Energy Accelerator Research Organization (KEK),
and also in part by the Grants-in-Aid of the Ministry of 
Education (Nos. 09304029, 10640246, 10640248, 10740107, 10740125,
11640294, 11740162).  K-I.I is supported by the JSPS Research
Fellowship.

%%%%%%%%%%%%%%%%%%%%%%%%%%%%%%%%%%%%%%%%%%%%%%%%%%%%%%%%%%%%%%%%%%%%%%

%%%%%%%%%%%%%%%%%%%%%%%%%%%%%%%%%%%%%%%%%%%%%%%%%%%%%%%%%%%%%%%%%%%%%%
\end{document}